\documentstyle[aps,preprint]{revtex}
\begin{document}
\draft
\preprint{WIS -- 93/69/July -- PH}
\title{Resonant Tunneling and the Pauli Principle}
\author{S. A. Gurvitz$,^{(1),(2)}\;$ H. J. Lipkin$,^{(1),(3)}\;$
and Ya. S. Prager$^{(1)}$}
\address{$^{(1)}$Department of Nuclear Physics, Weizmann Institute of
         Science, Rehovot 76100, Israel\\
$^{(2)}$TRIUMF, Vancouver, B.C., Canada V6T\ 2A3\\
$^{(3)}$School of Physics and Astronomy,
        Raymond and Beverly Sacler Faculty of Exact Sciences,
        Tel-Aviv University, Tel-Aviv 69978, Israel}
\date{\today}
\maketitle
\begin{abstract}
A new method based on modified optical Bloch equations is
proposed for studying resonant tunneling in semiconductor
heterostructures. The method manifestly takes account of
electrons' statistics, which enables to investigate the
influence of the Pauli principle on resonant tunneling
in presence of inelastic scattering.
Being applied to evaluation of the resonant current in
semiconductor heterostructures, our approach predicts
considerable deviations from the one-electron picture.
\end{abstract}

\narrowtext

The impressive progress in microfabrication technology
has considerably enhanced the interest to
quantum transport through multilayered semiconductor
heterostructures \cite{Tunn1991}.
A good number of theoretical approaches
have been developed to describe resonant tunneling in these
devices. However, the problem has been mostly treated
within the one-electron picture.
The existing many-electron approaches (see, e.\ g., \cite{Frensley745,%
Chen295,Davies4603,Datta1347})
are very complicated, owing to which the influence of
many-body effects in resonant tunneling problems remains
insufficiently elucidated yet. One of them is the effect of the
Pauli exclusion principle,
which is expected to be especially important in presence of
inelastic scattering \cite{Datta1347,Landauer110}.

We propose here a novel method of handling resonant-tunneling
problems. This method is based on modified master equations suited
for problems, where ``classical'' (incoherent) and ``quantum''
(coherent) phenomena are in
interplay. Being technically simpler than the other approaches, it has
nevertheless a many-body nature and takes account of the Pauli
principle from the very beginning.

Let us consider a mesoscopic
``device'' consisting of coupled quantum wells. The device is
connected with two
separate electron reservoirs. We assume that the density of states
in the reservoirs is very high (continuum). However, each of the quantum
wells in the device contains only a small number of discrete energy levels.
For the sake of simplicity we consider the reservoirs at zero
temperature, where the carriers (electrons) occupy all the states
in the left reservoir up to the Fermi level $E_F^{(l)}$ and in
the right reservoir up to the level $E_F^{(r)}<E_F^{(l)}$. As a result,
a current is flowing through the device
from the left reservoir to the right one. The device is also coupled
to an additional phonon reservoir (also at zero temperature), which
enables inelastic transitions between the discrete levels
in the quantum wells. We assume that the system
is described by a tunneling Hamiltonian, i.e. only the couplings
between the adjacent wells are taken into account.

The general idea of our approach follows from the observation that the
electron transport in the system can be fully determined in terms
of the density submatrix of the device solely, without resorting
to the complicated density matrix of the entire system including
the reservoirs.
This density submatrix contains full information
about currents and particle accumulation, and the only problem
is to determine its time evolution. This problem can be solved by
using, with some modifications, the technique of NMR theory \cite{Abragam}
or the Bloch-Maxwell equations \cite{CohenTannoudji}.

Let us assume that the reservoirs' states remain unchanged
during the process. We designate the density
submatrix of the device as $\sigma_{ab}(t)$ in terms of all possible
electrons' states ($|a\rangle$, $|b\rangle$, $|c\rangle,\ldots$)
inside the device. For instance, $|a\rangle$ denotes the state
where there are no carriers inside the device; $|b\rangle$ --
one electron occupies the upper level in the first well, and so on.
In the spirit of Bloch-Maxwell equations \cite{CohenTannoudji}, we write
the equations for diagonal and off-diagonal matrix elements
of the density submatrix as
\begin{mathletters}
\label{a1}
\begin{eqnarray}
\dot{\sigma}_{aa} &=& -i [ {\cal{H}}, \sigma ] _{aa}
                      -\sigma_{aa}\sum_{n(\neq a)} \Gamma_{a
                      \rightarrow n} + \sum_{c(\neq a)} \sigma
                      _{cc}\Gamma_{c \rightarrow a}\;,
\label{a1a}\\
\dot{\sigma}_{ab} &=& -i [ {\cal{H}}, \sigma ] _{ab}
                      -\frac{1}{2}\sigma_{ab}\left ( \sum_{n(\neq a)}
                      \Gamma_{a \rightarrow n} + \sum_{n(\neq b)}
                      \Gamma_{b \rightarrow n} \right )\;,
\label{a1b}
\end{eqnarray}
\end{mathletters}
where ${\cal{H}}$ is the matrix of the ``internal'' (without reservoirs)
Hamiltonian, and $\Gamma_{a \rightarrow n}$ is the probability
per unit time for the system to make a transition from
the state $|a\rangle$ to
the state $|n\rangle$ of the device due to the tunneling to (or from) the
reservoirs, or due to interaction with the phonon bath.
The commutator in these equations
describes the coherent (``quantum'') motion of carriers inside
the device, and the remaining
terms describe the incoherent (``classical'')
motion of the carriers. By neglecting the commutator one obtains the
``classical'' kinetic master equations.

Now we can use our basic assumption that the system is described by a
tunneling Hamiltonian, which takes into account only
the couplings between the
nearest neighbors. Then we can rewrite Eqs.~(\ref{a1})
explicitly as
\begin{mathletters}
\label{a2}
\begin{eqnarray}
\dot\sigma_{aa} & = &
 i\sum_{b(\neq a)}\Omega_{ab}(\sigma_{ab}-\sigma_{ba})
-\sigma_{aa} \sum_{n(\neq a)}\Gamma_{a \rightarrow n} +
 \sum_{c(\neq a)} \sigma_{cc}\Gamma_{c \rightarrow a}\;,
 \label{a2a}\\
\dot\sigma_{ab} & = & i(E_b - E_a) \sigma_{ab} +
i\left (\sum_{b'(\neq b)}\sigma_{ab'}\Omega_{b'b}
-\sum_{a'(\neq a)}\Omega_{aa'}\sigma_{a'b}\right )
\nonumber\\
    &  & -\frac{1}{2}\left ( \sum_{n (\neq a)}\Gamma_{a \rightarrow n}
                  +\sum_{n (\neq b)}\Gamma_{b \rightarrow n}\right )
                   \sigma_{ab}\;,
\label{a2b}
\end{eqnarray}
\end{mathletters}
where $\Omega_{ij}$ denote the couplings between the levels
in adjacent wells, and $a_{ba}=a^*_{ab}$.

By solving Eqs.~(\ref{a2}), one can find the
diagonal elements of the density submatrix,
which determine the charge density of the corresponding
quantum states of the device. Then the
current flowing into the right reservoir is
\begin{equation}
J(t)=\sum_i\sigma_{ii}\Gamma_R^{(i)}\;,
\label{a3}
\end{equation}
where the sum is taken over only those states $|i\rangle$ that
contain electrons in the well adjacent to the
right reservoir;
$\Gamma_R^{(i)}$ is the partial width of the state $|i\rangle$
due to tunneling to the right reservoir\cite{bg}. The stationary current
$I$ is obtained from Eq.~(\ref{a3}) as $I=J(t\rightarrow\infty )$ and
does not depend on the initial state of the device \cite{bg,g}.

Now we give some examples of application of our method to resonant
tunneling problems. Although the method can be applied
to many different problems,
we concentrate in this letter on the Pauli
exclusion effects in resonant tunneling.
In all the examples we neglect
transitions to motion parallel to the layers,
by assuming that the scattering on impurities and layers irregularities
is small. This makes the problem essentially one-dimensional.
For the sake of simplicity we also neglect the electron spin effects,
assuming only that two electrons cannot occupy the same quantum state.

Let us start with the well-known case of penetration
through a double-barrier heterostructure with one discrete level
$E_1$ inside the well, Fig.~1(a).
In this case the device has only two possible states, namely:
$|a\rangle$ -- the level $E_1$ is empty,
and $|b\rangle$ -- the level $E_1$ is occupied.
There are two possible interstate transitions:
$|a\rangle\rightarrow |b\rangle$
via $\Gamma_L$ (an electron tunnels from the left reservoir), and
$|b\rangle\rightarrow |a\rangle$ via $\Gamma_R$ (an electron tunnels
to the right reservoir). Hence Eqs.~(\ref{a2a}) give
\begin{mathletters}
\label{a4}
\begin{eqnarray}
\dot\sigma_{aa} & = & -\Gamma_L\sigma_{aa}+\Gamma_R\sigma_{bb}\;,
\label{a4a}\\
\dot\sigma_{bb} & = & \Gamma_L\sigma_{aa}-\Gamma_R\sigma_{bb}\;,
\label{a4b}
\end{eqnarray}
\end{mathletters}
where $\Gamma_{L,R}$ are the partial widths of the level $E_1$ due to
tunneling to the left and the right reservoirs. Solving Eqs.~(\ref{a4})
with the initial condition either $\sigma_{aa}(0)=1$, $\sigma_{bb}=0$,
or $\sigma_{aa}(0)=0$, $\sigma_{bb}=1$, and using Eq.~(\ref{a3}),
we obtain for the resonant current $I$ the standard result
\begin{equation}
I=\frac{\Gamma_L\Gamma_R}{\Gamma_L+\Gamma_R}\;.
\label{a5}
\end{equation}
Notice that the coherent (quantum) effects in the resonant current
are absent here. As it follows from Eqs.~(\ref{a2}), such
effects can appear only due to coupling between discrete levels
in adjacent quantum wells, whereas in the considered example
there is only one quantum well.
Hence, the transport through this system is described by classical
master equations. It thus might be not surprising that ``coherent'' and
``sequential'' approaches to the resonant tunneling produce the same
answer in this case, \cite{Toombs257}.

Next we consider the resonant tunneling through double-well potential
structure shown in Fig.~1b. As in the previous case,
there are no inelastic transitions due to
interaction with the phonon reservoir, since each of the wells contains only
one level. However, the {\em coherent} tunneling
between the wells can take place in this structure.
Let us enumerate all possible electron states in the device.
There are just four of them:
$|a\rangle$ -- the levels $E_{1,2}$ are empty; $|b\rangle$ -- the
level $E_1$ is occupied; $|c\rangle$ -- the level $E_2$ is occupied;
$|d\rangle$ -- both the levels $E_1$ and $E_2$ are occupied.
Taking account of all possible transition between these states,
one obtains from Eqs.~(\ref{a2})
\begin{mathletters}
\label{a6}
\begin{eqnarray}
\dot\sigma_{aa} & = & -\Gamma_L\sigma_{aa}+\Gamma_R\sigma_{cc}\;,
\label{a6a}\\
\dot\sigma_{bb} & = & \Gamma_L\sigma_{aa}+\Gamma_R\sigma_{dd}+
i\Omega (\sigma_{bc}-\sigma_{cb})\;,
\label{a6b}\\
\dot\sigma_{cc} & = & -\Gamma_R\sigma_{cc}-\Gamma_L\sigma_{cc}+
i\Omega (\sigma_{cb}-\sigma_{bc})\;,
\label{a6c}\\
\dot\sigma_{dd} & = & -\Gamma_R\sigma_{dd}+\Gamma_L\sigma_{cc}\;,
\label{a6d}\\
\dot\sigma_{bc} & = & i(E_2-E_1)\sigma_{bc}+
i\Omega (\sigma_{bb}-\sigma_{cc})-\frac{1}{2}\left (\Gamma_L+\Gamma_R
\right )\sigma_{bc}\;.
\label{a6e}
\end{eqnarray}
\end{mathletters}
The total current $I$, according to Eq. (\ref{a3}), is
$I=(\bar\sigma_{cc}+\bar\sigma_{dd})\Gamma_R$, where
$\bar\sigma_{ii}\equiv\sigma_{ii}(t\rightarrow\infty )$.
Eqs. (\ref{a6}) can be easily solved, so that we finally
get for the resonant current
\begin{equation}
I=\left (\frac{\Gamma_L\Gamma_R}{\Gamma_L+\Gamma_R}\right )
\frac{\Omega^2}{\Omega^2+\Gamma_L\Gamma_R/4+\epsilon^2\Gamma_L\Gamma_R/
(\Gamma_L+\Gamma_R)^2}\;,
\label{a7}
\end{equation}
where $\epsilon =E_2-E_1$.
This result coincides with the one found in the framework of
one-electron approach \cite{g,sok}. At first sight,
such a coincidence looks surprising.
Indeed, our treatment explicitly excludes
the states when two or more electrons occupy the same level. On the
other hand, these states are implicitly present
within the one-particle picture. Nevertheless,
a more
detailed analysis shows that this coincidence is not accidental.
One can demonstrate that
in the {\em absence of inelastic scattering}
the Pauli forbidden configurations in the one-electron picture
do not contribute to the resonant current \cite{gp}.
However, in the presence of inelastic scattering, the influence of Pauli
forbidden terms is very essential.

Let us consider two other examples of the resonant tunneling, now
in presence of inelastic scattering. The first case is shown in
Fig.~2a.
It corresponds to the double-barrier structure discussed above,
but now the quantum well contains two levels,
the upper one $E_0$ and the lower one $E_1$. An
electron which tunnels from the left reservoir into the upper level
$E_0$ can either relax inelastically into the lower state with the
transition rate $\Gamma_{in}$, and then tunnel out into the right
reservoir, or tunnel out directly into the right reservoir.
There are four possible electron states of the device:
$|a\rangle$ -- the levels $E_{0,1}$ are empty; $|b\rangle$ -- the upper
level, $E_0$, is occupied; $|c\rangle$ -- the lower level,
$E_1$, is occupied;
$|d\rangle$ -- both levels $E_0$ and $E_1$ are occupied.
The possible interstate transitions are:
$|a \rangle \rightarrow |b \rangle$ via $\Gamma_{L}$, which is the
tunneling rate from the left reservoir;
$|b \rangle \rightarrow |a \rangle$ via $\Gamma_{R0}$, which is the
tunneling rate to the right reservoir from the upper level;
$|b \rangle \rightarrow |c \rangle$ via $\Gamma_{in}$;
$|c \rangle \rightarrow |a \rangle$ via $\Gamma_{R1}$
which is the tunneling rate to the right reservoir from the lower level;
$|c \rangle \rightarrow |d \rangle$ via $\Gamma_{L}$;
$|d \rangle \rightarrow |b \rangle$ via $\Gamma_{R1}$;
$|d \rangle \rightarrow |c \rangle$ via $\Gamma_{R0}$.
Accordingly, using Eqs.~(\ref{a2}) we get
\begin{mathletters}
\label{a8}
\begin{eqnarray}
\dot\sigma_{aa} & = & -\Gamma_L\sigma_{aa}+\Gamma_{R0}\sigma_{bb}+
\Gamma_{R1}\sigma_{cc}\;,
\label{a8a}\\
\dot\sigma_{bb} & = & \Gamma_{L}\sigma_{aa}-(\Gamma_{R0}+\Gamma_{in})
\sigma_{bb}+\Gamma_{R1}\sigma_{dd}\;,
\label{a8b}\\
\dot\sigma_{cc} & = & \Gamma_{in}\sigma_{bb}
-(\Gamma_L+\Gamma_{R1})\sigma_{cc}+\Gamma_{R0}\sigma_{dd}\;,
\label{a8c}\\
\dot\sigma_{dd} & = & \Gamma_L\sigma_{cc}-(\Gamma_{R0}+\Gamma_{R1})
\sigma_{dd}\;.
\label{a8d}
\end{eqnarray}
\end{mathletters}
The total current $I=I_{R0}+I_{R1}$ is given by Eq.~(\ref{a3}), where
$I_{R0}=f_0\Gamma_{R0}$ is the current
flowing to the right reservoir from the upper level, and
$I_{R1}=f_1\Gamma_{R1}$ is the current
flowing from the lower level (the ``inelastic current'').
Here $f_0=\bar\sigma_{bb}+\bar\sigma_{dd}$
and $f_1=\bar\sigma_{cc}+\bar\sigma_{dd}$ are the
electron densities on the upper and the lower levels, correspondingly.
After solution of Eqs.~(\ref{a8}) we obtain
\begin{equation}
I_{R0}=\frac{\Gamma_L\Gamma_{R0}(1+\alpha )}{\Gamma_T +
\alpha (\Gamma_L+\Gamma_{R0})}\;,\;\;\;\;\;
I_{R1}=\frac{\Gamma_L\Gamma_{in}}{\Gamma_T +
\alpha (\Gamma_L+\Gamma_{R0})}\;,
\label{a9}
\end{equation}
where $\Gamma_T =\Gamma_L+\Gamma_{R0}+\Gamma_{in}$ and
$\alpha =\Gamma_L\Gamma_{in}/[\Gamma_{R1}(\Gamma_L+\Gamma_{R0}+
\Gamma_{R1})]$.

This result is quite different from the stationary
resonant current $I^{(\mbox{\scriptsize one-el})}$ having been
found within the one-electron framework \cite{bg,lip},
\begin{equation}
I_{R0}^{(\mbox{\scriptsize
one-el})}=\frac{\Gamma_L\Gamma_{R0}}{\Gamma_T}\;,\;\;\;\;\;
I_{R1}^{(\mbox{\scriptsize
one-el})}=\frac{\Gamma_L\Gamma_{in}}{\Gamma}\;.
\label{a10}
\end{equation}
Notice that $\Gamma_{R1}$ does
not enter Eq.~(\ref{a10}), and therefore the stationary current
in the one-electron approach does not depend on subsequent processes
following the electron relaxation to the lower level.
This follows from the irreversibility of the
inelastic scattering, which disconnects
the inelastic flux from the initial channel \cite{bg,lip}.
Hence, it is very remarkable that the
$\Gamma_{R1}$--dependence of the resonant
current appears in Eq.~(\ref{a9}).
It means that, when the exclusion principle is taken into account,
the irreversibility does not totally disconnect the processes
taking place before and after the relaxation.

One can show \cite{bg} that the inelastic current
$I_{R1}^{(\mbox{\scriptsize one-el})}$
in the one-electron approach, Eq.~(\ref{a10}), can be represented
as $I_{R1}^{(\mbox{\scriptsize
one-el})}=f_0\Gamma_{in}$, where $f_0$ is the electron
density on the upper level. Therefore one may tempt
to account for
the exclusion principle by writing the inelastic current as
$I_{R1}=f_0(1-f_1)\Gamma_{in}$, where $f_1$ is the electron density
on the lower level, Eq.~(\ref{a3}). However, one can
easily check, by using
Eq.~(\ref{a9}), that this assumption does not hold. This example shows
explicitly that {\em ad hoc} introduction of the additional factor
$(1-f)$ cannot be correct in general for
taking account of the Pauli principle
in quantum transport \cite{Datta1347,Landauer110,Bonig9203}.

It is important to mention that the total
current $I=I_{R0}+I_{R1}$, Eq.~(\ref{a9}), always increases
with $\Gamma_{in}$. It is similar to the result of the one-electron
treatment, Eq.~(\ref{a10}). However, the account for the exclusion
principle leads to restriction of the total resonant
current $I$ when $\Gamma_{in}\rightarrow\infty$
(as can be easily seen from Eq.~(\ref{a9})).

The most spectacular manifestation of the Pauli exclusion principle
would take place in the resonant tunneling through
double-well structures, in presence of inelastic transitions.
As an example we consider
the system shown in Fig.~2b, where a resonant current flows due to
inelastic transitions from the upper to the lower levels
in the left well. We investigate the case
where the lower level in the left well, $E_1$, and
the level $E_2$ in the right well are aligned, i.e. $E_1=E_2$, so
the electrons may oscillate between the wells.
As a result, the exclusion principle generates non-trivial
quantum interference effects in the course of inelastic transitions
from the upper level $E_0$ to the lower one $E_1$ \cite{footnote}.

Enumerating all possible electron states
one can straightforwardly write down
the system of linear coupled equations, Eqs.~(\ref{a2}),
for all transitions in the device. We do not present
these equations here, since the procedure is exactly the same as
in all the previous examples.
Here we give only the final expression for the resonant
current, Eq.~(\ref{a3}), when the inelastic rate
$\Gamma_{in}\gg\Gamma_{L,R},\Omega$:
\begin{equation}
I=\frac{{\cal M}\Gamma_L\Gamma_R}{{\cal N}+{\cal P}\Gamma_{in}
/2\Omega^2}\;.
\label{a11}
\end{equation}
Here the coefficients ${\cal M}$, ${\cal N}$ and ${\cal P}$
are functions of the tunneling
widths, $\Gamma_{L,R}$,
and the coupling $\Omega$ between the aligned levels, namely
\begin{mathletters}
\label{a12}
\begin{eqnarray}
{\cal M} & = & \Omega^2(\Gamma_L^2+4\Gamma^2)+2\Gamma_L\Gamma_1
\Gamma^2\;,
\label{a12a}\\
{\cal N} & = & \Gamma_L(2\Omega^2+\Gamma_1\Gamma)(4\Gamma^2-
\Gamma_L\Gamma_R)
+\frac{\Gamma_L^2\Gamma_R}{4}\left (\Gamma_R^2-\Gamma_L^2+
\frac{\Gamma_1
\Gamma\Gamma_R^2}{\Omega^2}\right )\;,
\label{a12b}\\
{\cal P} & = & \Gamma_1\Gamma_L^2\Gamma_R(\Omega^2+\Gamma^2)\;,
\label{a12c}
\end{eqnarray}
\end{mathletters}
where $\Gamma =(\Gamma_L+\Gamma_R)/2$ and
$\Gamma_1=(2\Gamma_L+\Gamma_R)/2$.
It follows from Eq. (\ref{a11}) that,
contrary to expectations,
the resonant current decreases to zero when the inelastic rate
$\Gamma_{in}$ increases.
Such a strange behavior of the tunneling (inelastic) current
is a result of quantum interference effects due to the Pauli
principle. Indeed, since the tunneling time from
the left well to the right one is much larger than the
time of inelastic
transition from the upper level, an electron from the upper level
should have arrived the second well ``simultaneously''
with the previous electron.
It leads to an effective localization of electrons at the upper and
the lower levels in the left
well, and therefore to the decrease of the inelastic resonant current.

One can also consider a
configuration where the level $E_1$ in Fig.~2b is above the
bottom of the Fermi sea in the left reservoir. Then the total
resonant current has two components: the first one from the inelastic
transitions from the upper level $E_0$, and the second one
from the direct tunneling to the level $E_1$. In this case, too,
the first component interferes destructively with the second one.
As a result, the total resonant current increases when the first
component is ``switched off'' (i.e. when $E_F<E_0$).

All the discussed deviations from the single particle picture
can be verified in experiments with semiconductor quantum wells.
Yet, one should take into account that inelastic
transitions with subsequent rescattering on impurities, with
the momentum transfer to directions parallel to the layers,
may decrease the expected deviations.

\begin{figure}[h]
\caption{Resonant tunneling
through (a) double-barrier and (b) triple-barrier heterostructures
in absence of inelastic scattering.}
\label{fig1}
\end{figure}
\begin{figure}[h]
\caption{Resonant tunneling
through (a) double-barrier and (b) triple-barrier heterostructures
in presence of inelastic scattering.}
\label{fig2}
\end{figure}

\begin{references}
\bibitem{Tunn1991} {\em Resonant Tunneling in Semiconductor Physics
        and Applications}, edited by L. L. Chang, E. E. Mendez, and
        C. Tejedor (Plenum, New York, 1991).
\bibitem{Frensley745} W. R. Frensley, Rev.\ Mod.\ Phys.\ {\bf 62},
        745 (1990).
\bibitem{Chen295} L. Y. Chen, C. S. Ting, Int.\ J. Mod.\ Phys.\ B
        {\bf 6}, 295 (1992).
\bibitem{Davies4603} J. H. Davies, S. Hershfield, P. Hyldgaard, and
        J. W. Wilkins, Phys.\ Rev.\ B {\bf 47}, 4603 (1993).
\bibitem{Datta1347} S. Datta, Phys.\ Rev.\ B {\bf 45}, 1347 (1992).
\bibitem{Landauer110} R. Landauer, Phys.\ Scr.\ {\bf T42}, 110 (1992).
\bibitem{Abragam} A. Abragam, {\em The Principles of Nuclear
        Magnetism} (Cla\-rendon Press, Oxford, 1961).
\bibitem{CohenTannoudji} C. Cohen-Tannoudji, J. Dupont-Roc, and
        G. Grynberg, {\em Atom-Photon Interactions: Basic Processes
        and Applications} (Wiley, New York, 1992).
\bibitem{bg} I. Bar-Joseph and S. A. Gurvitz, Phys. Rev. B {\bf 44}, 3332
(1991).
\bibitem{g} S. A. Gurvitz, Phys.\ Rev.\ B {\bf 44}, 11\ 924 (1991).
\bibitem{Toombs257} G. A. Toombs and F. W. Sheard, in
        {\em Electronic Properties of Multilayers and
        Low-Dimensional Semiconductor Structures}, edited by
        J. M. Chamberlain {\em et al}. (Plenum, New York, 1990), p.257.
\bibitem{sok} D. Sokolovski, Phys.\ Lett.\ A {\bf 132}, 381 (1988).
\bibitem{gp} Two different proofs of this statement were suggested
recently by Landauer \cite{Landauer110}. However, each of them rests on
substantial additional assumptions. In fact, one can give
a proof free of such
shortcomings, S. A. Gurvitz and Ya. S. Prager, unpublished.
\bibitem{lip} H. J. Lipkin, Phys.\ Rev.\ B {\bf 46}, 15\,534 (1992).
\bibitem{Bonig9203} L. B\"{o}nig and K. Sch\"{o}nhammer,
        Phys.\ Rev.\ B {\bf 47}, 9203 (1992).
\bibitem{footnote} The one-electron treatment produces the same result
for the inelastic resonant current, as in the previous example,
namely $I^{(\mbox{\scriptsize
one-el})}=\Gamma_L\Gamma_{in}/(\Gamma_L+\Gamma_{in})$. It
follows from the disconnection of the
inelastic flux from the initial channel, when the exclusion principle
is disregarded.
\end{references}
\end{document}